\documentclass[manuscript]{aastex}

\def\arcsec{$''$}

\slugcomment{To be submitted to the Astrophysical Journal}

\shorttitle{Photoionization Models}
\shortauthors{Mel\'endez et al.}

\begin{document}

\title{Uncovering the  Spectral Energy Distribution  in Active Galaxies Using High Ionization Mid-infrared Emission Lines}


\author{M. Mel\'endez\altaffilmark{1}, S.B. Kraemer\altaffilmark{2}, K. A. Weaver}
\affil{NASA Goddard Space Flight Center, Greenbelt, MD, 20771}

\author{R. F. Mushotzky}
\affil{Astronomy Department, University of Maryland, College Park, MD}


\altaffiltext{1}{Department of Physics and Astronomy, The Johns Hopkins University, Baltimore, MD, 21218}
\altaffiltext{2}{Institute for Astrophysics and Computational Sciences, Department of Physics, The Catholic University of America,Washington, DC 20064}

\begin{abstract}

The shape of the spectral energy distribution of active galaxies in the EUV--soft 
X-ray band (13.6~eV to 1~keV) is uncertain because  obscuration by dust and gas can hamper our 
view of the continuum.  To investigate the shape of the spectral energy distribution 
 in this energy band, we have generated a set of photoionization models which reproduce the small 
dispersion found in correlations between high-ionization mid-infrared emission 
lines in a sample of hard X-ray selected AGN. Our calculations show that a  
broken power-law continuum model is sufficient to reproduce the 
[Ne~V]$_{14.32 \micron}$/[Ne~III], [Ne~V]$_{24.32 \micron}$/[O~IV]$_{25.89 \micron}$ 
and [O~IV]$_{25.89 \micron}$/[Ne~III] ratios, and does not require the addition of 
a ``big bump" EUV model component. We constrain the  EUV--soft X-ray slope, 
$\alpha_i$, to be between 1.5 -- 2.0 and derive a best fit of  $\alpha_i \sim 1.9$ 
for Seyfert~1 galaxies, consistent with previous studies of intermediate redshift quasars. If we assume  a blue bump model, most sources in our sample have 
derived temperatures between  $T_{BB}=10^{5.18}$~K to  $10^{5.7}$~K, suggesting that the peak of 
this component spans a large range of energies extending from  
$\sim \lambda 600$~\AA~ to $\lambda 1900$~\AA. In this case, the best fitting peak 
energy that matches the mid-infrared line ratios of Seyfert~1 galaxies occurs 
between  $\sim \lambda 700$--$\lambda1000$~\AA. Despite the fact that our results do not rule out the presence of an EUV bump, 
we conclude that our power-law  model produces enough photons with energies $>$ 4~Ry to generate 
the observed amount of mid-infrared emission in our sample of BAT AGN. 
\end{abstract}

\section{Introduction}

Active galactic nuclei (AGN) produce an extraordinary amount of energy which is  
thought to be primarily thermal emission from an optically thick accretion disk fueling 
a supermassive black hole \citep[e.g.,][]{2004ApJ...613..682P}. Accretion can produce a characteristic  feature in the 
ultraviolet continuum of most AGN known as the ``big blue bump" \citep[bbb,][]{1978Natur.272..706S,1982ApJ...254...22M} 
and it has been argued that its energy peak lies around the Lyman limit 
\citep[e.g.,][]{1997ApJ...475..469Z,2002ApJ...565..773T,2005ApJ...619...41S}. The problem 
is that in AGN the soft X-ray--extreme-ultraviolet spectral energy distribution (SED)  can be 
hidden because of  dust  and gas,
 thus making a direct view of the full energy distribution impossible. A common approach 
is to study high redshift quasars where the rest-frame extreme-ultraviolet (EUV, $\lambda \sim$ 100--912~\AA) can be observed 
from the ground. However, for the local universe at z~$<$~0.05 we need to rely 
on space-borne telescopes and/or other indirect methods to uncover the 
shape of the SED in this particular range of energies.

 Within the paradigm of the unified model of AGN \citep[e.g.,][]{1993ARA&A..31..473A} 
both Type~1 and Type~II AGN are intrinsically the same, with their observed 
difference being due to the visibility of the central engine. An 
optically-thick toroidal structure surrounds the central source, located at 
distances just beyond the broad line region (BLR). Emission that arises further 
out in the narrow line region (NLR), hundreds of parsecs in extent, is expected 
to be fairly unobstructed and isotropic, i.e., independent of our viewing 
angle. However, absorption, dust extinction and star formation may hamper 
the use of soft X-ray, optical and infrared observables on these 
larger-than-torus scales \citep[e.g.,][]{1994ApJ...436..586M,2010arXiv1011.5993K}.

One way to bypass the above limitations is to 
focus on high-ionization (${\rm E > 40~eV}$), mid-infrared emission lines. 
To be produced, they require photons with energies between $\sim$40-100eV 
(EUV--Soft X-ray), and so we can assume that they are primarily associated with the AGN, in other words,  they are less contaminated 
by star formation processes. Also, they are less sensitive to dust extinction than optical lines. In this 
regard, it is possible to fit these observed lines via photoionization models  in order to constrain the AGN continuum in this energy band \cite[e.g.,][]{1999ApJ...512..204A,2000ApJ...536..710A}.  Among the mid-infrared properties of Seyfert galaxies, the correlations 
between high-ionization emission lines, [Ne~III]~$\lambda$15.56~\micron, 
[O~IV]~$\lambda$25.89~\micron~   and [Ne~V]~$\lambda$$\lambda$14.32/24.32~\micron, 
are strikingly tight, with a dispersion of less than $\sim 0.2$ dex, both in flux and 
luminosity \citep[see ][]{2007ApJ...655L..73G,2009ApJ...691.1501D,2009MNRAS.398.1165G,2008ApJ...689...95M,2010ApJ...716.1151W,2010ApJ...725.2270P}. The [Ne~V] emission, given the 
high energy require to be produce (${\rm E >}$98.9~eV), has been  unambiguously identified  
with the presence  of an AGN \citep[see][]{2007ApJ...663L...9S,2008ApJ...678..686A,2009ApJS..184..230B,2010ApJ...716.1151W}, therefore, the strong correlation between this  
emission  and the lines produced from transitions of [Ne~III] (${\rm E >}$ 41.0~eV) and [O~IV] (${\rm E >}$ 54.9~eV) suggest that 
these lines are produced primarily by the AGN. Moreover, \cite{2008ApJ...682...94M}  probed the utility of high-ionization 
mid-infrared lines as an indicator of AGN power. They found a tight 
correlation, with a dispersion of $\sim 0.3~{\rm dex}$, in Seyfert 1 galaxies, 
between the [O~IV]  and the X-ray 14-195~keV continuum luminosities. A weaker 
correlation was found for Seyfert 2 galaxies, which we believe is due to the  
effect of Compton Scattering in the 14-195~keV band in some Seyfert 2 galaxies. This result has since been 
confirmed in similar studies  
\citep[e.g.,][]{2009ApJ...700.1878R,2009ApJ...698..623D,2010ApJ...716.1151W}.

Our aim is to recover the hidden portion of the spectral energy distribution 
of AGN between the EUV and the soft X-ray (13.6~eV to 1~keV).  To do this we  have performed extensive photoionization calculations to characterize 
the small dispersion found in the correlations between high-ionization, mid-infrared emission-line ratios. We compare our calculations with emission
line ratios from {\it Spitzer}/IRS spectra of the {\it Swift} 
Burst Alert Telescope (BAT) AGN sample\footnote{The  {\it Swift}/BAT covers the whole sky at 
(1--3)$\times 10^{-11}{\rm ergs~cm^{-2}~s^{-1}}$ and represents a complete 
sample of active galaxies including Compton-thin AGN that were missed from 
previous X-ray surveys in the 2-10 keV band because of their high column 
densities ($N_{H}\sim 10^{24}\ \mathrm{cm}^{-2}$ ).} of the local universe, presented in \cite{2010ApJ...716.1151W}.  All of our calculations 
have been carried-out by  the photoionization code CLOUDY, version 08.00, 
last described by \cite{1998PASP..110..761F}.

\section{The Models}

\begin{figure}
\epsscale{0.9}
\plotone{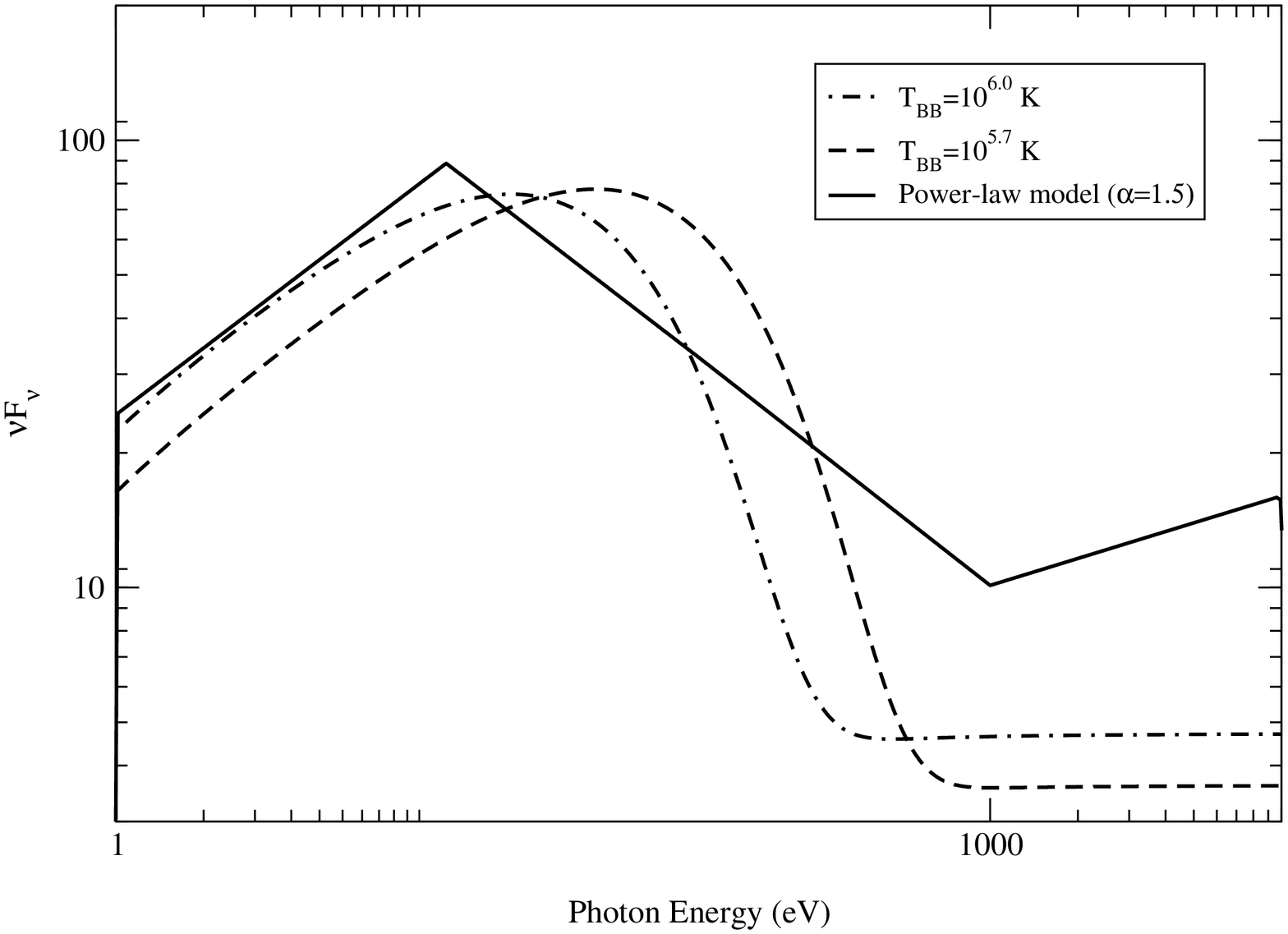}
\caption{ Comparison between the different continuum spectral energy distribution considered in the present work. For the sake of comparison, for the ``blue bump"  model we present  two  values for  the UV bump cutoff temperature $10^{5.70}$~K and $10^{6.0}$~K. Again, for comparison purpose, we used a ``power-law" model with a  value for the EUV/Soft X-ray index  of  $\alpha_i=1.5$. The three continua shown are normalized to have the same integrated  luminosity between 1~Ryd$\leq h\nu \leq 7.354\times 10^6$~Ryd, however, because we are focusing  on the modeling of emission line ratios, instead of of absolute line fluxes, the chosen normalization is arbitrary.   \label{fig1} }
\end{figure}

We have generated two sets of photoionization models to study the intrinsic 
SED of the ionizing continuum. The first assumes a broken power law, as 
used  by \cite{2008ApJ...682...94M} and  similar to that suggested for  NGC~5548 and NGC~4151 \citep{1998ApJ...499..719K,2000ApJ...531..278K},  of the form $F_\nu \propto \nu^{-\alpha}$, 
with $\alpha {\rm =0.5}$ below 13.6~eV, $\alpha_i$ from 13.6~eV to 1~keV and 
0.8 at higher energies. To examine specifically the EUV/soft X-ray
band, we varied $\alpha_i$ from 1.0 to 2.5.  Our second model is a 
parametrization of the AGN continuum that combines an X-ray power law and
a ``blue bump" component, in the form: 
 
\begin{equation}
f_\nu=\nu^{\alpha_{uv}}\exp{\left( -h\nu/kT_{BB}\right)}\exp{\left( -kT_{IR}/h\nu\right )} + a\nu^{\alpha_{x}},
\end{equation}   
where $\alpha_{uv}$ is the low-energy slope of the Big Blue Bump continuum, $T_{BB}$ is the UV bump cut-off temperature and  $kT_{IR}$ is the 
infrared cut-off temperature of the big bump, corresponding to 9.1~$\micron$. For the X-ray power-law, $\alpha_{x}$ is the slope of the X-ray 
component and the coefficient $a$ is adjusted to produce the correct  UV to X-ray slope \citep{1979ApJ...234L...9T},  $\alpha_{ox}$, as defined by
\begin{equation}
\frac{f_{\nu}(2 {\rm keV})}{f_{\nu}(2500{\rm \AA})}=403.3^{\alpha_{ox}}.
\end{equation}
For the ``big bump" model we adopted a value for the UV to X-ray spectral slope of $\alpha_{ox}=-1.4$ \citep{1981ApJ...245..357Z}, 
in good agreement with recent studies of Type~I AGN  \citep{2010A&A...512A..34L}. We  fixed  the low-energy slope of the big blue bump continuum to 
$\alpha_{uv}=-0.5$ \citep[e.g.,][]{1993AJ....106..417F}. For the X-ray power-law, we adopted  a value of $\alpha_{x}=-1$.  This parametrization is similar to 
that of \cite{1987ApJ...323..456M}  and \cite{1997ApJS..108..401K}. Finally, for this  model we considered five  values for  the UV bump cutoff temperature, from $T_{BB}=10^{5.18}$~K to  $10^{6.0}$~K. Figure~\ref{fig1} compares  the SED for our  different photoionization models.

To investigate the physical conditions in the emission line regions for  the [\ion{O}{4}], [\ion{Ne}{3}] and [\ion{Ne}{5}] lines, we started with a simple,   single-zone, constant density model. The logs of abundances of Oxygen and Neon relative to H by 
number are  -3.31 and -4.0, respectively, where the O abundance is  from \cite{2001ApJ...556L..63A} and the Ne abundance is from \cite{2001AIPC..598...23H}.  We considered 
hydrogen column densities of $10^{21}{\rm cm^{-2}}$ and $10^{22}{\rm cm^{-2}}$. Despite the fact that dust is likely mixed in with the emission-line gas in the NLRs 
of Seyfert galaxies \citep[e.g.,][]{1986ApJ...307..478K,1993ApJ...404L..51N}, the 
tightness of the mid-infrared correlation suggest that extinction in the mid-infrared, 
on average, is not an important bias in our sample of X-ray selected AGN \citep{2010ApJ...716.1151W}.

We generated  a grid of photoionization models varying   the  total hydrogen number density, ($n_h$), and the ionization parameter, $U$, where the ionization parameter  $U$ is defined as \cite[see][]{2006agna.book.....O}:
\begin{equation}
U=\frac{1}{4\pi R^2cn_H}\int^\infty_{\nu_o}\frac{L_\nu}{h\nu}d\nu=\frac{Q(H)}{4\pi R^2cn_H},
\label{u}
\end{equation}
where R is the  distance to the cloud, c is the speed of light  and $Q(H)$ is the flux of ionizing photons. In the next section we will compare these models  with the line ratios observed from the BAT AGN sample.

\section{Comparison with Observations}

\subsection{Density in the Emission Line Regions}

The use of emission lines that are close in ionization potential but  different critical densities can  be used as a density diagnostic of photoionized 
plasmas \citep[e.g.,][]{2006agna.book.....O}. In particular, the mid-infrared [Ne~V] ratio, [Ne~V]$_{14.32 \micron}$/[Ne~V]$_{24.32\micron}$, has been used as a  
density diagnostic in active galaxies \citep[e.g.,][]{2007ApJ...664...71D,2010ApJ...709.1257T} with the caveat that a wavelength-dependent mid-infrared extinction 
can affect this ratio, as suggested by the number of Seyfert galaxies that have [Ne~V] ratios below 
the theoretical low density limit \citep[e.g.,][]{2010ApJ...709.1257T,2010ApJ...716.1151W,2010ApJ...725.2270P}.  However, the 
mid-infrared extinction curve derived by \cite{2001ApJ...548..296W} and \cite{2001ApJ...554..778L}  \citep[e.g, see Figure~16 in][]{2001ApJ...554..778L} fails to 
predict the required dust extinction at ${\rm A_{14.32\micron}/N_H}$ and ${\rm A_{24.32\micron}/N_H}$ to explain the observed [Ne~V] ratios that fall below the low
density limit in the BAT sample \citep[][]{2010ApJ...716.1151W}. 
Therefore, since the relation between the [Ne~V] ratio and extinction is not fully understood, we  
minimized the possible effect of reddening toward the NLR by 
focusing on the [Ne~V] ratio observed in Seyfert~1 galaxies as detailed  multi-wavelength studies on Seyfert galaxies suggest more extinction 
of emission from the NLR in Seyfert~2 galaxies than in Seyfert~1s \citep[e.g.,][]{2008ApJ...682...94M,2011ApJ...727..130K}.  
Following the  work on the mid-infrared properties of the  BAT sample of AGN \citep{2010ApJ...716.1151W} we found an average neon ratio observed in  
Seyfert~1 galaxies of [Ne~V]$_{14.32 \micron}$/[Ne~V]$_{24.32\micron}=0.95\pm0.28$ (see Table~\ref{mean}).

\begin{figure}
\epsscale{0.9}
\plotone{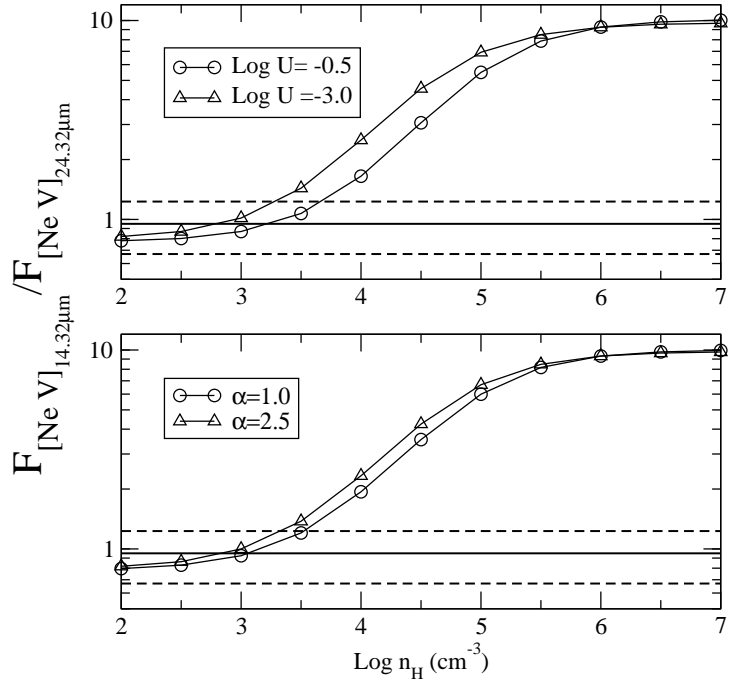}
\caption{ Upper panel. The [Ne~V] ratio versus the hydrogen number density, $n_H$ for  the ``power-law" model  with $\alpha_i=1.5$ for different ionization parameters, $\log U=-0.5$ and $\log U=-3.0$.  Lower panel. The [Ne~V] ratio versus the hydrogen number density, $n_H$, for  the ``power-law" model for different spectral slopes  with a fix ionization parameter, $\log U=-1.5$. The solid horizontal line represents the observed average ratio  and  the 1-$\sigma$ ({\it dashed lines}) found for Seyfert~1 galaxies. \label{fig2} }
\end{figure}

We used the {\bf optimize}\footnote{For a full description on the optimization methods see CLOUDY documentation, Hazy~1, at http://www.nublado.org/} command in CLOUDY 
to vary $n_h$ and $U$ to reproduce the observed neon ratio. From this, and by using our ``power-law" model with $\alpha_i=1.5$ for a total hydrogen column density of $n_H=10^{21}{\rm cm^{-2}}$,
we obtained a value of $n_h=10^{2.96}~{\rm (cm^{-3})}$ and $\log U=-1.37$.
For  our ``blue bump" model, assuming a cutoff temperature of $T_{BB}=10^{5.7}$~K such that the energy of the UV bump 
peaked at $\sim 22$~eV \citep[e.g.,][]{1997ApJS..108..401K}, we obtained a value of $ n_h=10^{2.91}~{\rm (cm^{-3})}$ and $\log U=-1.51$. 
For a total hydrogen column density of $N_H=10^{22}~{\rm cm^{-2}}$ we obtained a value of $n_h=10^{2.95}~{\rm (cm^{-3})}$ and $\log U=-1.31$ for our power-law models 
and, from our ``blue bump" model, assuming a cutoff temperature of $T_{BB}=10^{5.7}$~K, we obtained a value of $ n_h=10^{2.91}~{\rm (cm^{-3})}$ and $\log U=-1.30$. 
In Figure~\ref{fig2} we present the neon ratio versus the total hydrogen number density for these models for a range of ionization parameters and  EUV/Soft X-ray  slopes. 
Figure~\ref{fig2} shows that, close to the observed values, the neon ratio is not strongly sensitive to the SED of the ionizing radiation. 

From the different models and continuum approximations we obtained a total hydrogen number density of $n_h\approx 10^{3}~{\rm (cm^{-3})}$. 
One must note that this value represents the density only of the [Ne~V] emission-line gas. However, this density is in agreement with the values 
found from the photoionization modeling of the [O~IV] emitting region presented in \cite{2008ApJ...682...94M},  suggesting that the [Ne~V] and [O~IV] 
emitting regions are essentially co-located. On the other hand, this density is smaller than that found 
by \cite{2005MNRAS.358.1043B} for the [O~III] emitting region in their single-zone approximation, i.e., $n_e\sim 10^{5.85}~{\rm (cm^{-3})}$. 
However, they  found that the observed  [O~III]~$\lambda$5007, [O~III]~$\lambda$4363 and H$\beta$ lines can also be fitted using two-zone model comprised of a a dense 
($n_e\sim 10^{7}~{\rm cm^{-3}}$) inner zone, dominated by the  [O~III]~$\lambda$4363 emission, and a low density, more extended zone ($n_e\sim 10^{3}~{\rm cm^{-3}}$),
where most of the [O~III]~$\lambda$5007 arises. Our density value is similar to that used in their low-density zone, although the densities  in  their two-zone approximation 
are based on fixed assumed  values, whereas our density is derived from the fit to the observed [Ne~V] ratio.

\subsection{The EUV--Soft X-ray slope}

\begin{figure}
\epsscale{0.9}
\plotone{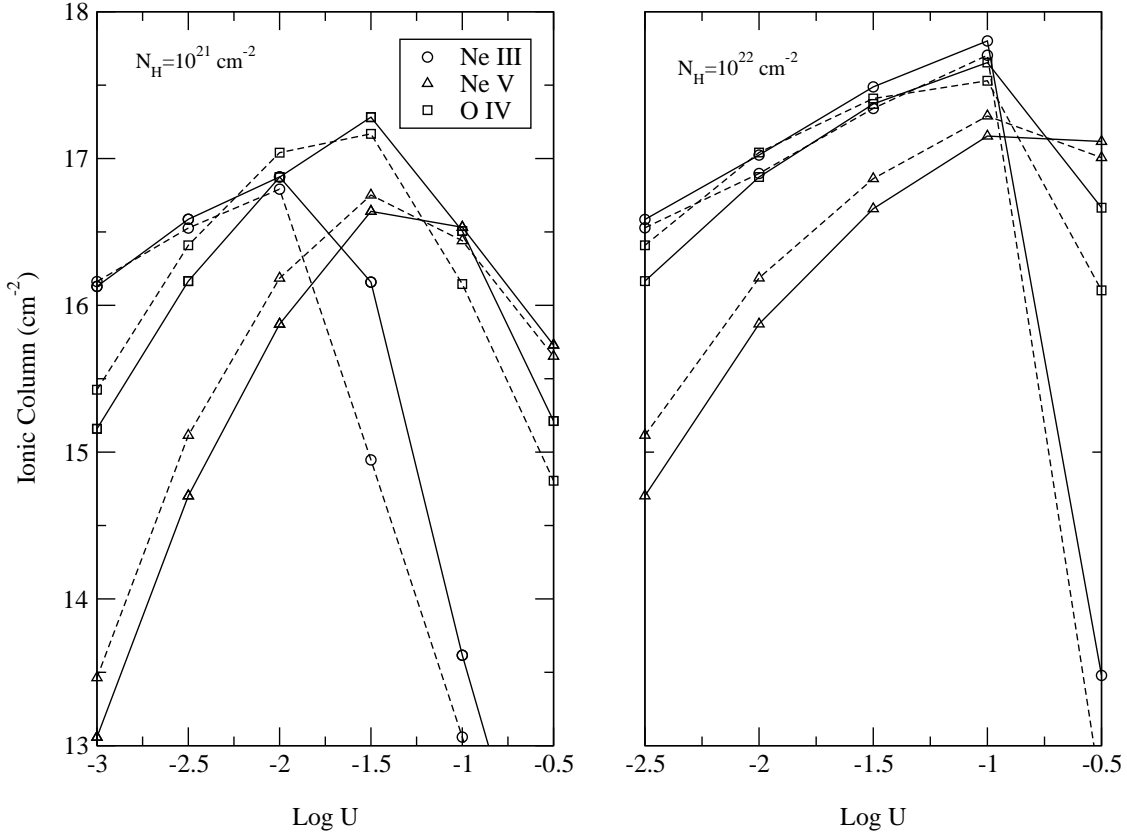}
\caption{Predicted ionic column densities for Ne~III ({\it circles}), Ne~V ({\it triangles}) and O~IV ({\it squares}), for two set of models, as a function of ionization parameter U, and two values  of the column density, $10^{21}{\rm cm^{-2}}$ and $10^{22}{\rm cm^{-2}}$, left and right panels, respectively. The solid line represents the ``power-law" model  with $\alpha_i=1.5$ and the dashed line represents 
the ``big bump" parametrization with the peak of the UV bump at $\approx 44$~eV, or alternatively a  UV bump cutoff temperature of $10^6$~K. \label{fig3} }
\end{figure}

\begin{figure}
\epsscale{0.9}
\plotone{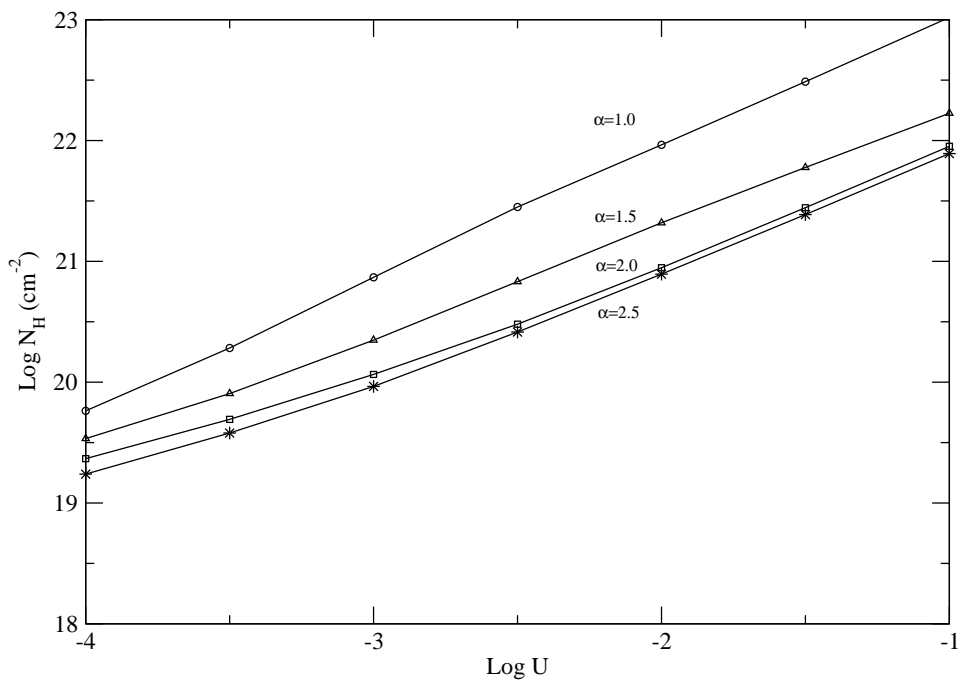}
\caption{Comparison for the  radiation bounded models between the total Hydrogen column density and ionization parameter  for the ``power-law" model for different values of the EUV/Soft X-ray index, $\alpha_i$. One must note from this comparison that, within the range of ionization parameters that reproduce the observed ratios, our radiation bounded models yield to Hydrogen column densities higher than that found in typical NLR conditions, $\log N_H=21~{\rm (cm^{-2})}$ \citep[e.g.,][]{2000ApJ...531..278K} \label{htotal} }
\end{figure}

\begin{figure}
\epsscale{0.9}
\plotone{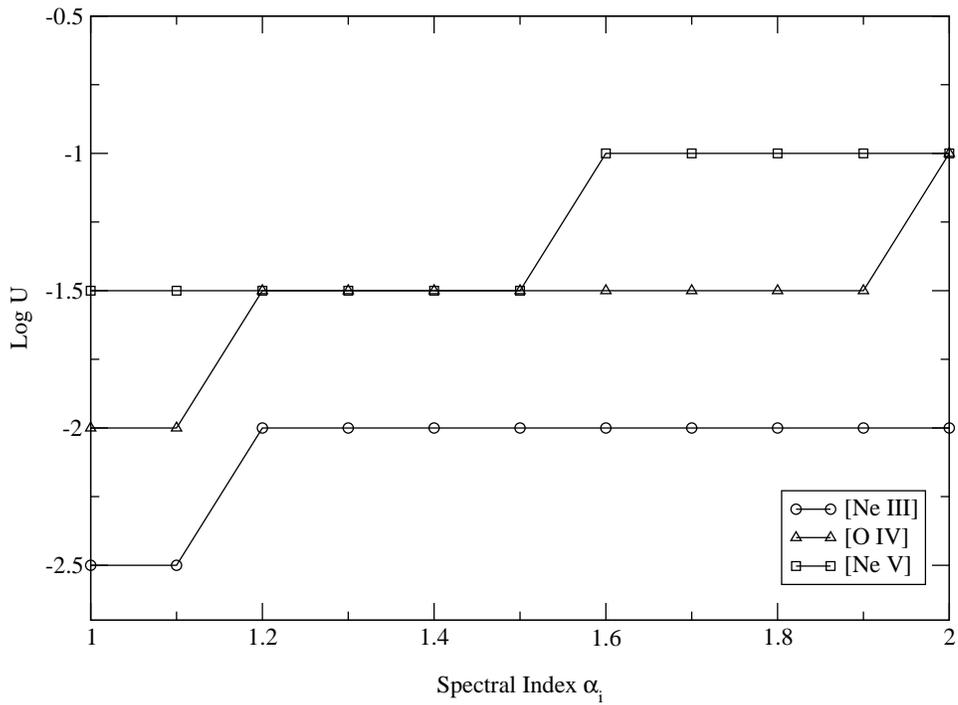}
\caption{ A comparison between the ionization parameter associated with the peak of the ionic column density versus the EUV/Soft X-ray  spectral slopes for the ``power-law" model. From this comparison is clear that the peak on the ionic column density depends on the shape of the ionizing continuum. \label{ionic} }
\end{figure}

Following the results from the previous section, and in order to  keep the number of free parameters to a minimum,
 we used  the observed [Ne~V] ratios to fix the hydrogen density in our photoionization models to $n_h= 10^3~{\rm (cm^{-3})}$. 
Next, in order to  maximize  each of the mid-infrared emission lines, we proceeded to find the range in ionization parameter where 
the predicted ionic column densities for the Ne~III, Ne~V and O~IV peak (see Figure~\ref{fig3}). Results show that the ionization parameter falls in the range
$-2.0 < \log U < -1.5$ for ${\rm N_H}=10^{21}{\rm cm^{-2}}$, while it is in a higher ionization range, $\log U \sim -1.0$, for ${\rm N_H}=10^{22}{\rm cm^{-2}}$.
One must note that the peak of the ionic column density depends on the SED (see Figure~\ref{ionic}). Also, because we are interested in the shape of the ionizing continuum,
we expect to achieve the clearest constraints from radiation-bounded models \footnote{Specifically, in a radiation-bounded model, all of the incident radiation has been absorbed inside the cloud, i.e., 
the cloud is optically thick to the ionizing radiation.}. The set of radiation-bounded models was obtained by initially assuming a hydrogen column density 
of $ {\rm N_H}=10^{22}{\rm cm^{-2}}$. In Figure~\ref{htotal} we compared the total Hydrogen density as a function of ionization parameter and UV slope for our radiation bounded, ``power-law" models. 
It is clear that a column density of ${\rm N_H}=10^{22}{\rm cm^{-2}}$ is enough to generate an optically thick, radiation-bounded cloud for the set of ionization parameters 
and hydrogen density derived from the previous considerations, except for $\alpha_i=1.0$ where we had to extend the column density to ${\rm N_H}=10^{23}{\rm cm^{-2}}$. 
Using this set of ionization parameters, and $n_h=10^3~{\rm cm^{-3}}$,  we compared our model predictions with the observed [Ne~V]$_{14.32 \micron}$/[Ne~III], [Ne~V]$_{24.32 \micron}$/[O~IV]
and [O~IV]/[Ne~III] ratios from the BAT sample. Overall, these results suggest that the [O~IV] and [Ne~V] come from a higher ionization state and lower density regions than 
that suggested by the optical [O~III] lines \citep[][]{2005MNRAS.358.1043B,2008ApJ...682...94M}. 

\subsection{The Power-Law model}

\begin{figure}
\epsscale{0.9}
\plotone{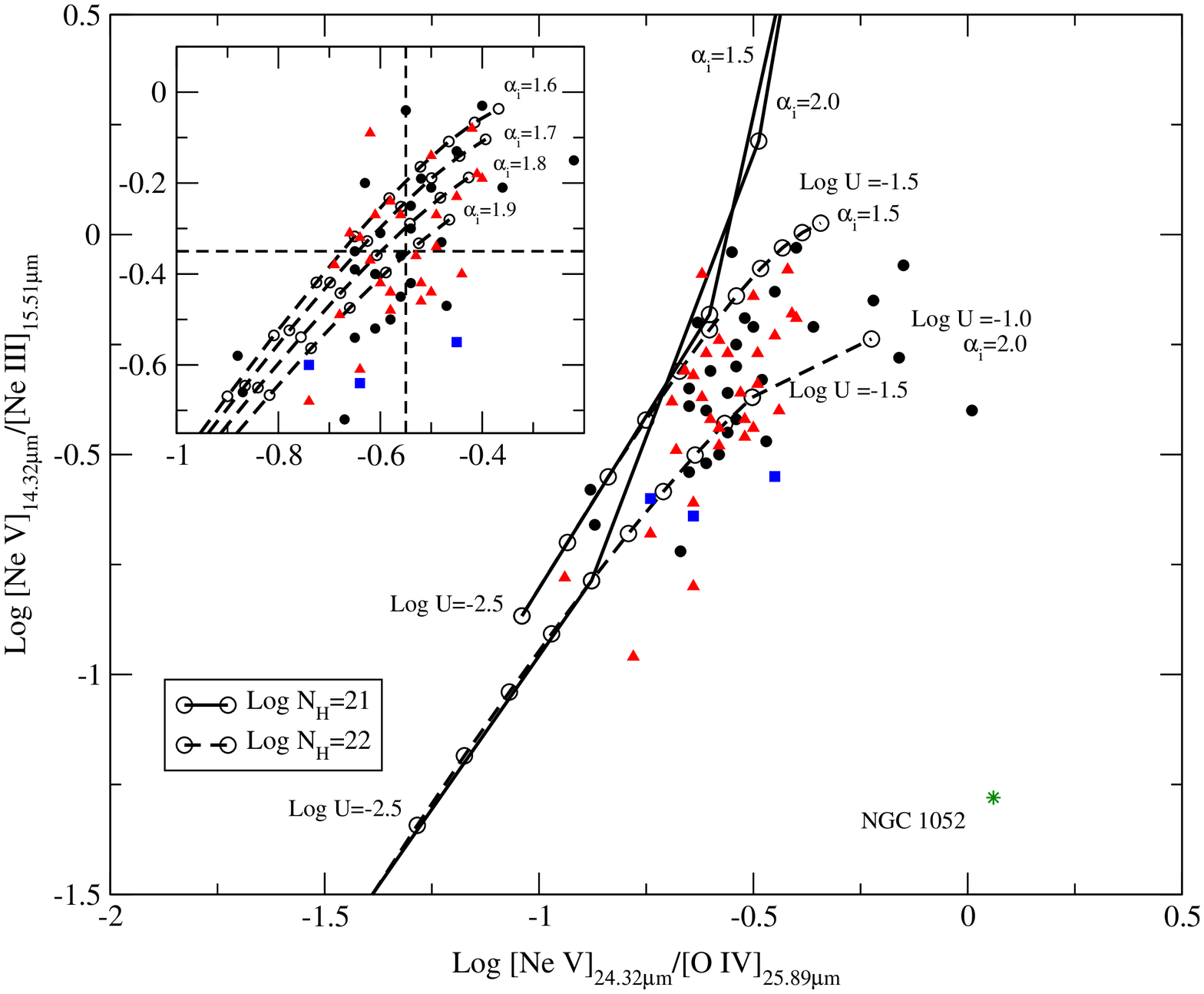}
\caption{Comparison between the predicted and observed  [Ne~V]$_{14.32 \micron}$/[Ne~III] and  [Ne~V]$_{24.32 \micron}$/[O~IV] ratios. Seyfert~1 galaxies are presented as black circles, Seyfert~2 galaxies are red triangles, blue squares represent the newly detected BAT AGN and green stars are LINERs.  We present  two set of  ``power-law" models  corresponding to $\log N_H=21~{\rm (cm^{-2})}$ ({\it solid} lines)  and $\log N_H=22~{\rm (cm^{-2})}$ ({\it dashed} lines) for a range of   $\alpha_i=1.5-2.0$. For the set of models presented in this comparison we  show  a range of ionization parameters of $-2.5 < \log U < -1.5$, in steps of 0.5~dex and 0.1~dex for $\log N_H=21$ and $\log N_H=22$, respectively. All these models corresponds to a hydrogen number density of $n_h=10^3$~(cm$^3$). \label{fig4} }
\end{figure}

As we mentioned before, high-ionization mid-infrared ratios can be used to investigate different excitation mechanisms, thus 
different shapes of the ionizing continuum. They provide information about the relative flux density of ionizing photons with energies 
between $\sim$40--100~eV.  In Figure~\ref{fig4} we compared the observed [Ne~V]$_{14.32 \micron}$/[Ne~III] and  [Ne~V]$_{24.32 \micron}$/[O~IV] ratios 
found in the BAT sample (see Table~\ref{mean}) against those  predicted by our photoionization calculations. Because of 
the proximity of wavelengths and equal aperture extraction for the emission lines in 
each ratio\footnote{The [Ne~V]$_{14.32 \micron}$ and [Ne~III] fluxes were extracted from the  {\it Spitzer}/IRS in the short-high order (SH, 4.7\arcsec $\times$ 11.3\arcsec). 
On the other hand, the [Ne~V]$_{24.32 \micron}$ and [O~IV] emission were extracted from  the {\it Spitzer}/IRS in the long-high order (LH, 11.1\arcsec $\times$ 22.3\arcsec)}, 
this comparison is unaffected by  dust extinction and aperture effects.  
Figure~\ref{fig4} shows  two sets of  ``power-law" models,  corresponding to $N_H=10^{21}~{\rm cm^{-2}}$ and $N_H=10^{22}~{\rm cm^{-2}}$, for a range of the EUV/soft X-ray
from $\alpha_i=1.5-2.0$. We show that, for a column density of $N_H=10^{21}~{\rm cm^{-2}}$, the cloud is optically thin to the ionizing radiation,
as the models underpredict the amount of [Ne~III] emission. Specifically, this result describes a matter-bounded component, in which some fraction of the ionizing radiation 
passes through the gas unabsorbed. Overall, a softer ionizing continuum will result in a lower [Ne~V]/[Ne~III] ratio, because there are fewer photons capable of ionizing Ne$^{+3}$.

\begin{figure}
\epsscale{0.9}
\plotone{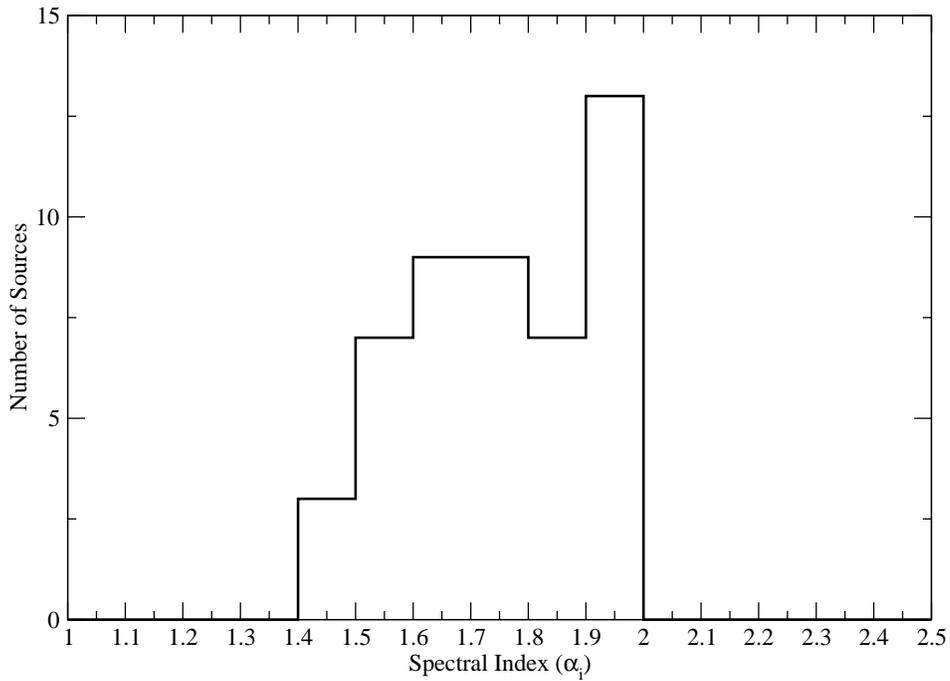}
\caption{Distribution of EUV/Soft X-ray slopes, $\alpha_1$,  derived from our analysis of the entire BAT sample. One must note that the spectral index was derived empirically from the position of each source in Figure~\ref{fig4}, therefore, each slope has an $\alpha_i\pm$1  error. For any given source located between two consecutive  power-law models, e.g., different spectral slope, we selected the flatter index as the empirical estimate.  \label{hist} }
\end{figure}

Figure~\ref{fig4} shows that, independent of the ionization parameter,  most of the observed ratios are matched by models with EUV/Soft X-ray spectral indices of $1.5 < \alpha_i < 2.0$. Using the 95$\%$ 
confidence interval for the actual mean for the [Ne~V]$_{14.32 \micron}$/[Ne~III] and  [Ne~V]$_{24.32 \micron}$/[O~IV] ratios observed in Seyfert~1 galaxies in our sample,
we found a smaller range of spectral indices, i.e.,  $1.7 < \alpha_i < 2.0$, with $\alpha_i \approx 1.9$  providing the best fit for the median observed values. A histogram of 
the EUV/Soft X-ray spectral indices, derived empirically from 
the position of each source in Figure~\ref{fig4}, is presented in Figure~\ref{hist}.  From their analysis of  Hubble Space Telescope ({\it HST}) Faint Object Spectrograph (FOS)
observations of a sample of 101 quasars at z $>$ 0.33, \cite{1997ApJ...475..469Z} found that the power-law index in the EUV between 350 and 1050~\AA~ is $\alpha=1.96\pm0.15$.
They also found that, for a sub-sample of 60 radio-loud sources, the EUV spectral index is $\alpha\approx2.2$, while for a sub-sample of 41 radio-quiet it is $\alpha\approx1.8$. 
As shown in Figure~\ref{fig4}, most of the BAT AGN can be fitted with an EUV/Soft X-ray spectral index of $\alpha < 2.0$, consistent with the radio-quiet nature of most Seyfert galaxies.

From their analysis of {\it HST} spectra of a  sample of 184 quasars at z $>$ 0.33, \cite{2002ApJ...565..773T} found a slightly harder EUV continuum than \cite{1997ApJ...475..469Z}. 
This sample included observations with the {\it HST} FOS, Goddard High Resolution Spectrograph and available Space Telescope Imaging Spectrograph (STIS) data. For their 
complete sample, they found a spectral index of $\alpha = 1.76 \pm 0.12$ between 500 and 1200~\AA~. For a subsample of radio-loud objects, they  found $\alpha =  1.96 \pm 0.12$, while
for a radio-quiet subset, they found $\alpha = 1.57 \pm 0.17$. These values are roughly consistent  with our range of EUV/Soft X-ray indices and the spectral index associated with the 
median values for the Seyfert~1 galaxies found in our sample, $\alpha \approx 1.9$. Again, this supports the idea that a single power-law can represent the EUV/soft X-ray continuum 
in our sample of BAT AGN. One must note that star formation contamination in the [Ne~III] emission can bias the EUV/Soft X-ray slope towards a steeper power-law index. In other words,  
star formation contamination in the [Ne~III] emission will result in a lower [Ne~V]$_{14.32 \micron}$/[Ne~III] ratio, which could be wrongly associated with a softer EUV continuum. 
However, the tight correlation found in the [Ne~III]--BAT relationship suggests that, on average, there is no strong enhancement due to star formation in the [Ne~III] emission 
in our sample, although we cannot rule out relatively minor effects in individual objects. Despite the fact that our spectral index has been determined over a relatively small range
of energy ($\sim$40--100~eV), these results support the validity of our method, given the agreement with values derived from different spectral analyses, covering a wider range of energies.

\begin{figure}
\epsscale{0.9}
\plotone{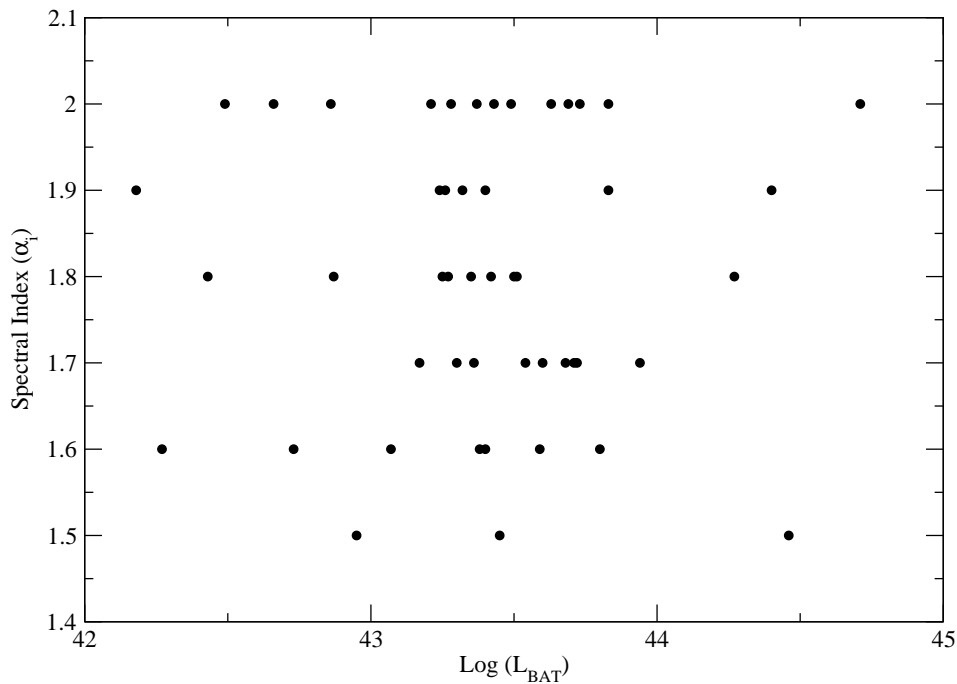}
\caption{Comparison between the spectral index, $\alpha_1$, and the 14--195~keV luminosities where it can be seen that there is  no dependence between the EUV/Soft X-ray slope and the 14--195~keV luminosity in our sample. One must note that the spectral index was derived empirically from the position of each source in Figure~\ref{fig4}, therefore, each slope has an $\alpha_i\pm$1  error. For any given source located between two consecutive  power-law models, e.g., different spectral slope, we selected the flatter index as the empirical estimate. \label{lum} }
\end{figure}

On the other hand, analysis of the Far Ultraviolet Spectroscopic Explorer ({\it FUSE}) archival data by \cite{2004ApJ...615..135S} revels a significantly harder composite spectral energy
distribution, $\alpha=0.56^{+0.38}_{-0.28}$, than the spectral indices discussed above and no evidence of a UV spectral index break. One must note that, in their sample, some individual 
sources do show a spectral break in the far-ultraviolet ($\lambda \sim$ 910--2000~\AA).  It should be noted that this is an heterogeneous sample of AGN, which were observed with {\it FUSE} 
because they were known to 
be bright in the UV. The authors reported a correlation between the spectral slope and AGN luminosity; in other words, low-luminosity AGN tend to show a harder ionizing continua. 
They suggested that the harder spectral index they derived is due to the relatively large fraction of low luminosity AGN in the {\it FUSE} sample\footnote{The median value for their 
{\it FUSE} sample is $\log \lambda L_{1100}=45.0~{\rm (ergs/s)}$ versus $\log \lambda L_{1100}=45.9~{\rm (ergs/s)}$ for the {\it HST} sample of active galaxies presented 
by \cite{2002ApJ...565..773T}}. In contrast, we found no evidence of such flatter indices in our sample. Specifically, in Figure~\ref{lum} we compare our derived values for the spectral index and 
the 14--195~keV luminosities, from which it can be seen that there is no dependence between the EUV/Soft X-ray slope and luminosity in the BAT sample. Furthermore,
\cite{2005ApJ...619...41S} found that most objects in their {\it FUSE} sample 
exhibit a spectral break around $1100$~\AA~, in agreement with previous {\it HST} results \citep{1997ApJ...475..469Z,2002ApJ...565..773T}. Their analysis is based on a heterogeneous 
sample of 17 low-redshift AGN (z$<0.5$), with quasi-simultaneous spectrophotometry including {\it FUSE}, {\it HST} and Kitt Peak National Observatory observations covering a rest wavelength 
from $900-9000$~\AA. Given the fact that \cite{2004ApJ...615..135S} looked for possible systematic errors that could bias their results, the source of this discrepancy is unclear. 
While it is possible that the spectral break is luminosity dependent, another possible explanation is that intrinsic reddening, if not fully corrected for, may lead to an incorrect 
determination of the spectral break and continuum shape \citep{2005ApJ...619...41S}.

\begin{figure}
\epsscale{0.9}
\plotone{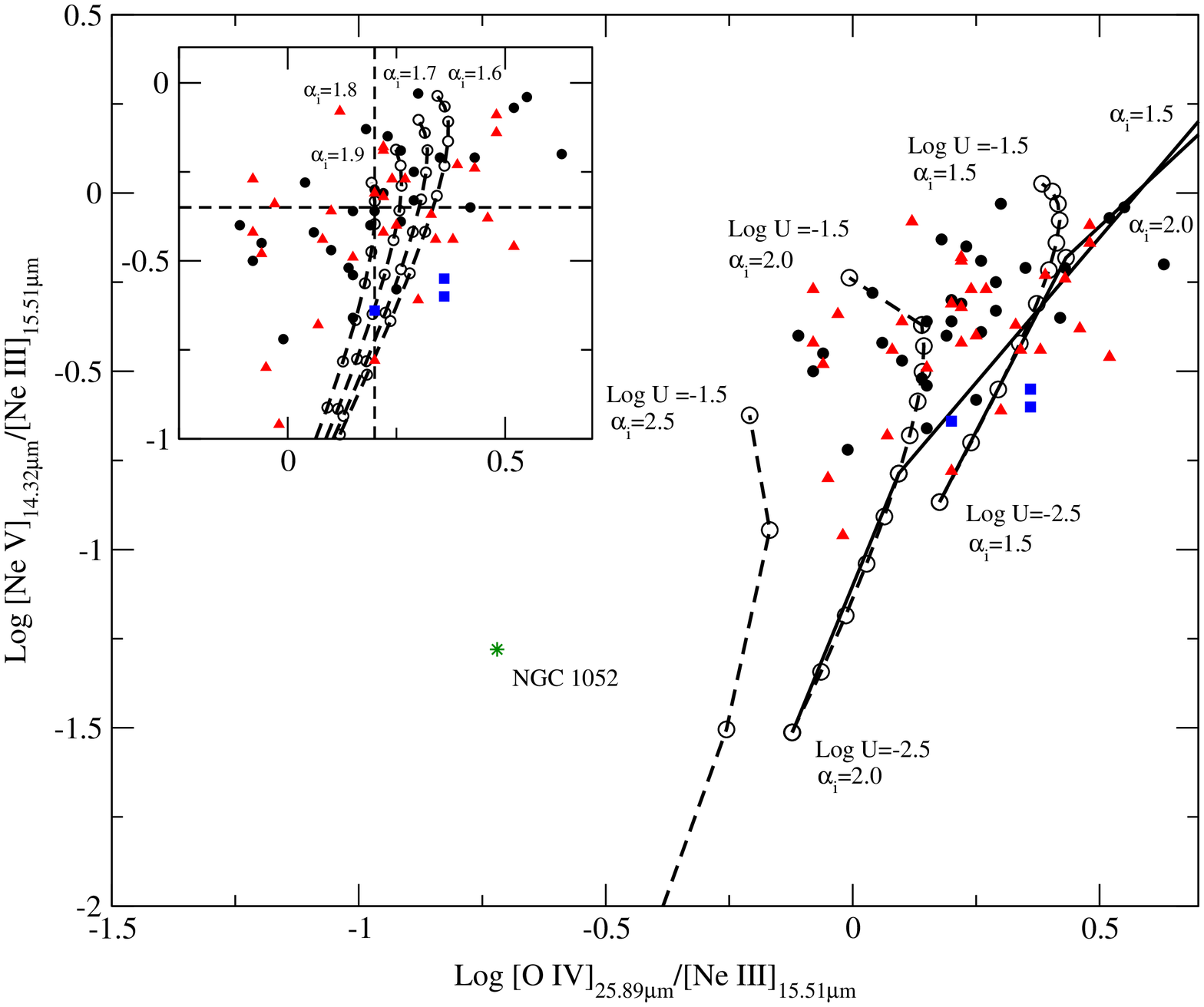}
\caption{Comparison between the predicted and observed  [Ne~V]$_{14.32 \micron}$/[Ne~III] and  [O~IV]/[Ne~III] ratios. Seyfert~1 galaxies are presented as black circles, Seyfert~2 galaxies are red triangles, blue squares represent the newly detected BAT AGN and green stars are LINERs.  We present  two set of  ``power-law" models  corresponding to $\log N_H=21~{\rm (cm^{-2})}$ ({\it solid} lines) and $\log N_H=22~{\rm (cm^{-2})}$ ({\it dashed} lines) for a range of   $\alpha_i=1.5-2.5$. For the set of models presented in this comparison we  show  a range of ionization parameters of $-2.5 < \log U < -1.0$, in steps of 0.5~dex and 0.1~dex for $\log N_H=21$ and $\log N_H=22$, respectively. All these models corresponds to a hydrogen number density of $n_h=10^3$~(cm$^3$). \label{fig5} }
\end{figure}

Figure~\ref{fig5} shows the predicted and observed [Ne~V]$_{14.32 \micron}$/[Ne~III] and [O~IV]/[Ne~III] ratios. Contrary to the ratios presented in  Figure~\ref{fig4}, the [O~IV]/[Ne~III]
ratio could be  susceptible to reddening  and  aperture effects. In this regard, these effects will have different manifestations in the observed [O~IV]/[Ne~III] ratios. For example,
the [O~IV]/[Ne~III] ratio will have larger values if extinction is stronger for the shorter-wavelength line. On the other hand, aperture effects will tend to decrease the observed ratio.  
As seen in Figure~\ref{fig5}, when using the  [O~IV]/[Ne~III] ratio, we recovered the same range for the EUV/soft X-ray slope as those derived from the [Ne~V]$_{14.32 \micron}$/[Ne~III] and
[Ne~V]$_{24.32 \micron}$/[O~IV]  ratios (see Figure~\ref{fig4}). In other words, the [O~IV]/[Ne~III] ratios are fit by the range of spectral indices found from the other ratios, 
suggesting that reddening  and aperture effects are not dominant (or, perhaps, cancel each other out) in the BAT sample of AGN, in agreement with the discussion presented
 in \cite{2010ApJ...716.1151W}.

\subsection{The AGN Parametrization}

\begin{figure}
\epsscale{0.9}
\plotone{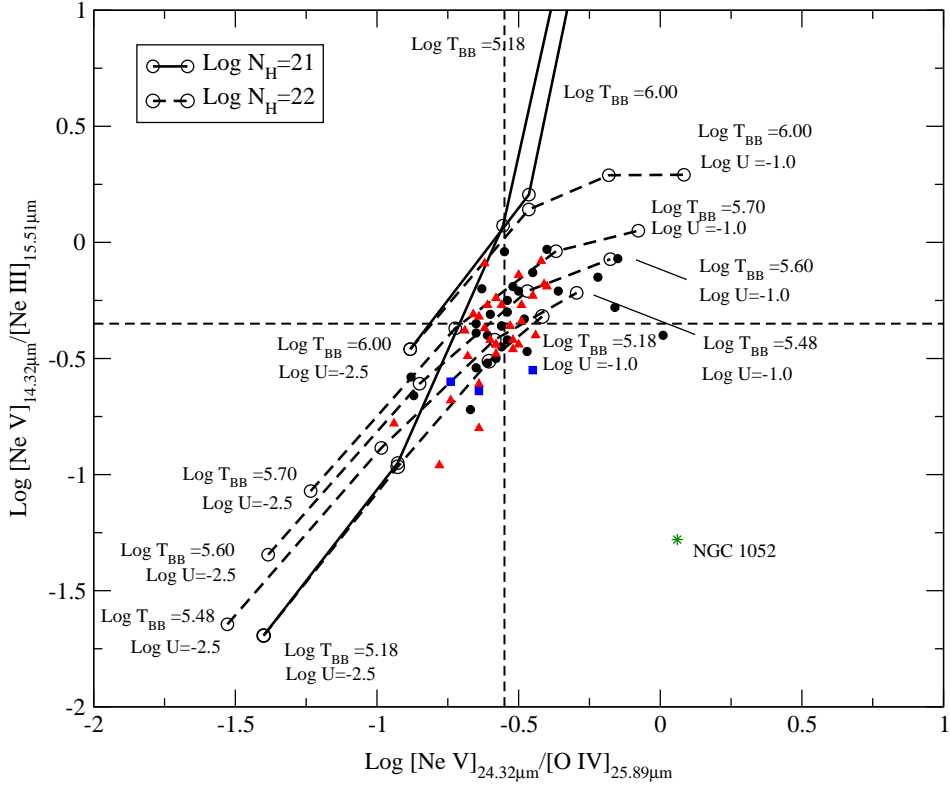}
\caption{Comparison between the predicted and observed   [Ne~V]$_{14.32 \micron}$/[Ne~III] and  [Ne~V]$_{24.32 \micron}$/[O~IV] ratios. Seyfert~1 galaxies are presented as black circles, Seyfert~2 galaxies are red triangles, blue squares represent the newly detected BAT AGN and green stars are LINERs.  We present  two set of  ``blue bump" models  corresponding to $\log N_H=21~{\rm (cm^{-2})}$ ({\it solid} lines) and $\log N_H=22~{\rm (cm^{-2})}$ ({\it dashed} lines) for five  values of  the UV bump cutoff temperature between $T_{BB}=10^{5.18}$~K  and $10^{6.0}$~K. For the set of models presented in this comparison we  show  a range of ionization parameters of $-2.5 < \log U < -1.0$, in steps of 0.5~dex for both $\log N_H=21$ and $\log N_H=22$. All these models corresponds to a hydrogen number density of $n_h=10^3$~(cm$^3$). \label{fig6} }
\end{figure}

\begin{figure}
\epsscale{0.9}
\plotone{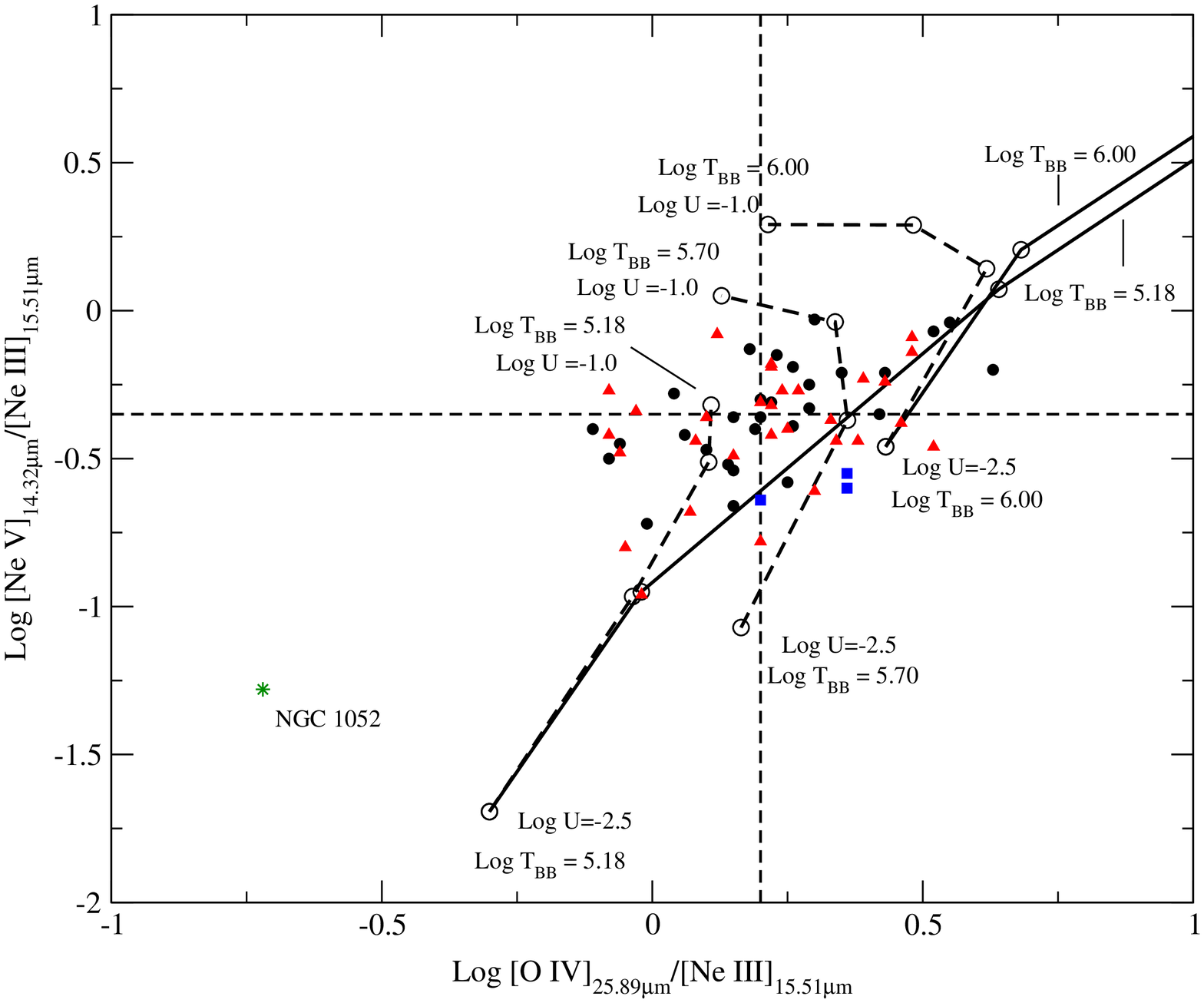}
\caption{Comparison between the predicted and observed  [Ne~V]$_{14.32 \micron}$/[Ne~III] and  [O~IV]/[Ne~III] ratios. Seyfert~1 galaxies are presented as black circles, Seyfert~2 galaxies are red triangles, blue squares represent the newly detected BAT AGN and green stars are LINERs. For the sake of simplicity,  we present  two set of  ``blue bump" models  corresponding to $\log N_H=21~{\rm (cm^{-2})}$ ({\it solid} lines) and $\log N_H=22~{\rm (cm^{-2})}$ ({\it dashed} lines)  only for three values of  the UV bump cutoff temperature, $T_{BB}=10^{5.18}$~K, $10^{5.70}$~K and $10^{6.0}$~K. For the set of models presented in this comparison we  show  a range of ionization parameters of $-2.5 < \log U < -1.0$, in steps of 0.5~dex for both $\log N_H=21$ and $\log N_H=22$. All these models corresponds to a hydrogen number density of $ n_h=10^3$~(cm$^3$). \label{fig7} }
\end{figure}

Figures~\ref{fig6}--\ref{fig7} show predictions for the [Ne~V]$_{14.32 \micron}$/[Ne~III], [Ne~V]$_{24.32 \micron}$/[O~IV] and [O~IV]/[Ne~III] ratios from a set of models generated 
with the parametrization of the AGN continuum including a UV bump, with cutoff temperatures from  
$T_{BB}=10^{5.18}$~K to  $10^{6.0}$~K, and column densities of $N_H=10^{21}~{\rm cm^{-2}}$ and $N_H=10^{22}~{\rm cm^{-2}}$.
Once again, at $N_H=10^{21}{\rm cm^{-2}}$ the cloud is optically thin to the ionizing radiation, thus underpredicts the [Ne~III] emission. It is clear that the AGN parametrization
with a  UV ``bump" temperature higher than  $10^{5.7}$~K  fails to predict most of the observed ratios, even in our radiation-bounded models. We find that the emission-line ratios for
a majority of sources can be fit by models with UV bump temperatures of $T_{BB}=10^{5.18}$~K to  $10^{5.7}$~K, or, alternatively, peaks at $\sim 600--1900$~\AA. The best fit for the 
95$\%$ confidence interval for the actual mean occurs between  $\sim 700--1000$~\AA. These results are in agreement with the FUV break observed in the composite spectra of AGN, in the
range of 1000--1100~\AA~\citep[e.g.,][]{1997ApJ...475..469Z,2002ApJ...565..773T,2005ApJ...619...41S}.

This range of temperature is slightly lower than the typical temperatures used to characterize an  average quasars continuum when modeling the BLR \citep{1997ApJS..108..401K}. 
However, the chosen form for the parametrization of the AGN continuum presented in \cite{1997ApJS..108..401K} focused on luminous quasars while our sample is dominated by lower luminosity 
sources. If the ionizing continuum is from an accretion disk, one would expect more luminous objects to be hotter for the same black hole mass, and,  therefore, the  
lower range of temperature found in our sample is not unexpected. However, these results may also argue in favor of a different, intrinsic  SED photoionizing the BLR. This intrinsic 
SED may be characterize by a bbb with a higher ``blue bump" temperature, e.g.,  $T_{BB} > 10^{5.7}$~K, than that found from our modeling of the NLR emission. In this interpretation, 
material located between the BLR and the NLR may absorb the intrinsic UV component. However, from our calculations, we do not find any clear evidence of a modification of the SED by an intervening 
medium.

\section{Conclusions}

In order to investigate the shape of the spectral 
energy distribution in the range 13.6~eV to 1~keV, we have generated photoionization models to characterize the small 
dispersion found in correlations between high-ionization mid-infrared 
emission-line ratios in AGN  This is the band where 
dust extinction and the line of sight absorber can make it virtually impossible to have a direct view 
of the continuum  generated by the nuclear engine. Our set of radiation 
bounded models show that a  broken power-law continuum model is 
sufficient to reproduce the observed  mid-infrared   
[Ne~V]$_{14.32 \micron}$/[Ne~III], [Ne~V]$_{24.32 \micron}$/[O~IV]$_{25.89 \micron}$
 and [O~IV]$_{25.89 \micron}$/[Ne~III] ratios. From this we constrained the EUV/Soft X-ray spectral index, 
$\alpha_i$, to be between 1.5 -- 2.0, with $\alpha_i \approx 1.9$  
corresponding to the median value observed in Seyfert~1 galaxies.  Our result 
agrees with previous studies \citep[e.g.,][]{1997ApJ...475..469Z,2002ApJ...565..773T,2005ApJ...619...41S} and 
holds for the wide  range of BAT luminosities and  type I and II objects in the local universe, whereas  
previous studies of SED tend to focus on UV selected objects at a different redshift range. Despite the fact that our spectral index has been estimated
for a relatively small range of energy ($\sim$40--100~eV), these results probe  the validity of our  method to uncover the  AGN continuum in this energy band. 
From the comparison between the photoionization models and observations we derived empirical values for the EUV/Soft X-ray slope for each source and found no dependence 
between the EUV/Soft X-ray slope and the 14--195~keV luminosity in our sample.

We also find that our power-law model can produce enough 
photons at energies $>$4~Ry to generate the amount of mid-infrared line emission observed in 
the BAT AGN sample. However, this result does not rule out the presence of a bbb. When we do add a ``big bump" 
in the UV to an X-ray power-law, the majority of sources  required ``blue bump" 
temperatures of $T_{BB}=10^{5.18}$~K to  $10^{5.7}$~K,  suggesting that the peak of this component 
lies in a wide range of energies that can extend from  $\sim 600$\AA~ to 
$1900$~\AA. The best fit for the  95$\%$ confidence interval for the actual mean occurs between,  $\sim 700$--$1000$~\AA, in
 agreement with previous estimates of the spectral index break \citep[e.g.,][]{1997ApJ...475..469Z,2002ApJ...565..773T,2005ApJ...619...41S}.  
 This range in 
temperatures is lower than the typical UV bump peak assumed to characterize an average quasar continuum when modeling the BLR \citep{1997ApJS..108..401K}. However,  if the ionizing
 continuum is from an accretion disk, one expects more luminous objects to be hotter for the same black hole mass. Therefore,  the  lower range of temperature found in our sample is
 not unexpected.


\acknowledgments
We would like to thank our anonymous referee for suggestions that improved the paper. This research was supported by an appointment to the NASA Postdoctoral Program at the Goddard Space Flight Center, administered by Oak Ridge Associated Universities through a contract with NASA. This research has made use of NASA's Astrophysics Data System.

\clearpage
\bibliographystyle{apj}
\bibliography{ms} 

\clearpage

\clearpage

\clearpage

\clearpage

\clearpage

\clearpage

\clearpage

\clearpage

\clearpage

\clearpage

\clearpage

\clearpage

\begin{deluxetable}{lccccccc}
\tabletypesize{\scriptsize}
\tablewidth{0pt}
\tablecaption{Statistical Analysis Between Seyfert 1 and Seyfert 2 Galaxies}
\tablehead{ & \multicolumn{3}{c}{Seyfert 1}& \multicolumn{3}{c}{Seyfert 2}\\
 \cline{2-4} \cline{5-7}\\
 & \colhead{Measurements} && Standard & Measurements &&Standard &${\rm P_{K-S}}$\\
 &\colhead{Available} & Mean&Deviation &Available&Mean&Deviation& (\%)}
\startdata
${\rm [Ne~V]_{14.32\micron}/[Ne~V]_{14.32\micron}}$&  29   &  0.95      &   0.28        & 30   &   1.00     &   0.38   &   87.8    \\  
${\rm [Ne~V]_{14.32\micron}/[Ne~III]}$&  32   &  0.47      &   0.22        & 33   &   0.41     &   0.20   &   79.8    \\  
${\rm [Ne~V]_{24.32\micron}/[O~IV]}$&  31  &   0.34      &   0.19      &  30   &    0.28     &  0.06  &      57.0 \\
${\rm [O~IV]/[Ne~III]}$&  38  &    1.58     &  0.87     &   33  &    1.65    & 0.77     &    91.4   \\

\label{mean} 
\enddata
\tablecomments{The last column, ${\rm P_{K-S}}$,  represents the Kolmogorov-Smirnov (K-S) test null probability. Upper limits for the  [Ne~V] fluxes are not included. This table also includes information about the numbers of Seyfert~1 and Seyfert~2 galaxies, mean values and standard deviations of the mean for the measured quantities. For the ${\rm [Ne~V]_{14.32\micron}/[Ne~III]}$ ratios the  95$\%$ confidence interval for the actual mean is  0.3904 -- 0.5465 for Seyfert~1 galaxies, and 0.3417 thru 0.4831 for Seyfert~2's. For the ${\rm [Ne~V]_{24.32\micron}/[O~IV]}$ ratios the  95$\%$ confidence interval for the  actual mean is  0.2717 thru 0.4089 for Seyfert~1 galaxies and 0.2469 thru 0.2964 for Seyfert~2's. Finally, for the ${\rm [O~IV]/[Ne~III]}$ ratios the   95$\%$ confidence interval for the actual mean is 1.295 thru 1.867   for Seyfert~1 galaxies, and  1.375 thru 1.924  or Seyfert~2's.
}  
\end{deluxetable}

\end{document}